\documentclass[conference]{IEEEtran}
\IEEEoverridecommandlockouts
\usepackage{cite}
\usepackage{amsmath,amssymb,amsfonts}
\usepackage{algorithmic}
\usepackage{graphicx}
\usepackage{textcomp}
\usepackage{xcolor}
\usepackage{comment}

\def\BibTeX{{\rm B\kern-.05em{\sc i\kern-.025em b}\kern-.08em
    T\kern-.1667em\lower.7ex\hbox{E}\kern-.125emX}}

\begin{document}

\title{Collecting Influencers: a Comparative Study of Online Network Crawlers\\
\thanks{
The study was funded by RFBR according to research project 18-07-01211.
\copyright 2019 IEEE. Personal use of this material is permitted.  Permission from IEEE must be obtained for all other uses, in any current or future media, including reprinting/republishing this material for advertising or promotional purposes, creating new collective works, for resale or redistribution to servers or lists, or reuse of any copyrighted component of this work in other works.
}
}

\author{\IEEEauthorblockN{Mikhail Drobyshevskiy$^{1,2}$, Denis Aivazov$^{1,2}$, Denis Turdakov$^{1,3}$,\\
Alexander Yatskov$^{1}$, Maksim Varlamov$^{1}$, and Danil Shayhelislamov$^{2}$}
\IEEEauthorblockA{$^1$\textit{Ivannikov Institute for System Programming of the Russian Academy of Sciences, Moscow, Russia}}
\IEEEauthorblockA{$^2$\textit{Moscow Institute of Physics and Technology (State University), Moscow, Russia}}
\IEEEauthorblockA{$^3$\textit{Lomonosov Moscow State University, Moscow, Russia}\\
\{drobyshevsky,aivazov,turdakov,yatskov,varlamov\}@ispras.ru, shayhelislamov.ds@phystech.edu}
}

\maketitle

\begin{abstract}

Online network crawling tasks require a lot of efforts for the researchers to collect the data. One of them is identification of important nodes, which has many applications starting from viral marketing to the prevention of disease spread. Various crawling algorithms has been suggested but their efficiency is not studied well. In this paper we compared six known crawlers on the task of collecting the fraction of the most influential nodes of graph.

We analyzed crawlers behavior for four measures of node influence: node degree, k-coreness, betweenness centrality, and eccentricity. The experiments confirmed that greedy methods perform the best in many settings, but the cases exist when they are very inefficient.

\end{abstract}

\begin{IEEEkeywords}
Network crawling, network sampling, node influence
\end{IEEEkeywords}

\section{Introduction}

Today there is a growing interest in network data collecting from online sources. Social networks, such as Facebook and Twitter, often provide APIs which help researchers obtain the data. However, several challenges arise here. A large scale of real-world graphs requires a huge amount of resources to crawl them due to bandwidth limits, many sites impose query limitations, etc.
Goals of network sampling could be different. Beside collecting the whole graph itself, one is interested in sampling a representative subgraph to use it instead of the original one, estimating network parameters, or estimating node/edge attributes~\cite{ahmed2014network}.

In many applications it is not necessary to crawl all nodes of a network but its most influential nodes only. Efficient identification of top-$k$ influential nodes is important for detecting key persons in social networks, preventing disease spread, controlling computer worms, and so on.
%  *ref TBD*
Node influence is associated with its centrality measure in the graph. For instance, node with high betweenness centrality is likely to have a high impact on information spread in a social network or a contagion process in a biological network.

A good network crawler should discover the highest centrality nodes with a minimal number of steps. Usually, a significant fraction of target nodes can be collected in relatively few iterations, especially when the degree distribution is skewed. For example, 5\% of nodes being sampled via random walk, cover 80\% of the $k$ largest degree nodes~\cite{lim2011online}.

Literature contains a number of crawling algorithms, while their efficiency depends on multiple factors. The choice of seed node, presence of ``protected'' users (whose connections can be detected only by incoming links from other observed nodes)~\cite{ye2010crawling}, and network structure~\cite{areekijseree2018guidelines} significantly influence the result performance of each method.

% постановка задачи, гипотезы которые хотим проверить и кратко наши результаты 
In this paper we compare several popular crawlers on the task of collecting a target set of top-10\% influential nodes of the graph. The distinctive feature of our study is that we run a crawler until it collects the whole graph, while crawlers are usually analyzed under a limited budget of queries. Our main contributions are the following ones.
\begin{enumerate}
\item The decision on the choice of crawling method significantly depends on the given budget. We observed that while a graph is being crawled, the leading algorithm can change several times.
\item Greedy methods, like the one guided by the maximal observed degree (MOD), are better than others in collecting the fraction of nodes with highest degrees, highest k-coreness, and highest betweenness centrality. Nodes with the least eccentricity, in comparison to other centralities, are harder to find for existing crawlers.
\item We confirm that MOD is often the best choice for network crawling, but there exist cases, when it loses to all other algorithms. The same conclusion holds for influential nodes crawling task.
\end{enumerate}

% sectioning 
In the next section we formalize the task and describe our experimental methodology. Section~\ref{sec:experiments} is devoted to the experimental results and their explanation suggestions. Then we provide a brief overview of related works in section~\ref{sec:related} and, finally, give a conclusion in section~\ref{sec:conclusion}.

\section{Problem definition and methodology}

\subsection{Problem Definition}
\label{sec:problem}

In our work, we consider a static unobserved undirected network, represented with a graph $G(V,E)$, where $V$ is the set of nodes and $E$ is the set of edges.
A crawler starts with a seed node $v_{seed} \in V$. Two sets, initially empty, are defined and dynamically updated at each step. $V'_c \subseteq V$ is a set of already closed (queried) nodes. $V'_o \subseteq V$ is a set of observed, but not closed nodes. At each iteration, the crawler queries the next node~$v \in V'_o$, which becomes closed, and updates $V'_o$ with newly seen neighbours of~$v$ in~$G$. Sampled graph $S_i=(V', E')$ at iteration $i$ consists of all closed and observed nodes $V' = V'_c \cup V'_o$ and all connections between them $E' \subseteq E$. We denote as $deg(v, S)$ the degree of node $v$ within graph $S$ (since $S$ is a sample of $V$, $deg(v, S)\le deg(v, V) $) and as $clust(v, S)$ its clustering coefficient.

\begin{table*}[h!]
\centering
\caption{Network crawling algorithms with a short description and computation complexity per one iteration.}
\begin{tabular}{|l|p{2.8cm}|p{10cm}|p{2.9cm}|}
\hline
\textbf{Name} &\textbf{Method} & \textbf{Node selection strategy} & \textbf{1 step complexity} \\ \hline \hline  
\textbf{RC} & Random Crawling & At each step, selects a random node from $V_o$ is selected. Does not depend on previously crawled node. & $O(\langle deg(v, G) \rangle)$ %считаем относительно |V|, |E|, $d_{avg}$
\\ \hline
\textbf{RW} & Random Walk & Random neighbour of previously crawled node %(until he could perform a query on node from $V_o$). Doesn't get stuck in communities. [1]
& $\gtrapprox O(\langle deg(v, G) \rangle)$ \\ \hline
\textbf{DFS} & Depth First Search & Traverses the graph in depth-first manner & $O(\langle deg(v, G) \rangle)$ \\ \hline
\textbf{BFS} & Breadth First Search & Traverses the graph in breadth-first manner & $O(\langle deg(v, G) \rangle)$ \\ \hline
\textbf{MOD} & Maximum Observed Degree & At each step selects a node from $V'_o$ with maximal degree
& $ O(\log |V'_o| \cdot \langle deg(v, S) \rangle) $ \\ \hline
\textbf{DE} & Densification-Expansion crawler & Switches between RW (expansion phase) and MOD analogue (Densification phase) strategies depending on statistics of the sampled graph~\cite{areekijseree2018guidelines}.
& $O(|V'_o| \cdot \langle deg(v, S)^2 \rangle) $

\\ \hline
%\hline
\end{tabular}
\label{tab:crawlers}
\end{table*}

\begin{table*}[h!]
\centering
\caption{Dataset of undirected networks. All parameters correspond to the giant component used in experiments.}
\begin{tabular}{|l|l|r|r|c|}
\hline  
\textbf{Name} & \textbf{Description} & \textbf{$|V|$} & \textbf{$|E|$} & \textbf{$\langle deg(v, G) \rangle$}
%& \textbf{ Q (modulation)}&\textbf{min and max ecc } 
\\ \hline \hline  
hamsterster & friendship graph of Hamsterster & 2\,000 & 16\,097 & 16
\\ \hline
DCAM & community subgraph from VKontakte & 2\,752 & 68\,741 & 50
\\ \hline
facebook & contains friendship data of Facebook users (2009) & 63\,392 & 816\,886 & 26
\\ \hline
slashdot & reply network of technology website Slashdot & 51\,083 & 131\,175 & 5.1
\\ \hline
github & membership network of the software development hosting site Github & 120\,865 & 439\,858 & 7.3
\\ \hline
dblp2010 & co-authorship network & 226\,413 & 716\,460 & 6.3
\\ \hline
\end{tabular}
\label{tab:graphs}
\end{table*}

The goal of the crawler is to cover a target set of most influential nodes $V^* \subseteq V$ as soon as possible. We consider four different measures of node influence: the degree, k-coreness, betweenness centrality, and eccentricity.
\begin{itemize}
\item Node \textit{degree} is the most straightforward measure of importance as the number of friends, subscribers, connections, citations, etc.

\item \textit{Betweenness centrality} characterizes how many paths in graph go through the node. High betweenness means high influence on information flows.

\item Node \textit{k-coreness} indicates that the node is a part of a connected subgraph where all nodes have degree at least~$k$.

\item Node \textit{eccentricity} measures the maximal distance to any other node in the graph. The lower the eccentricity, the faster information/disease spreads from the node to the rest of the graph.
\end{itemize}

\begin{figure}[h!]
\includegraphics[scale=0.13]{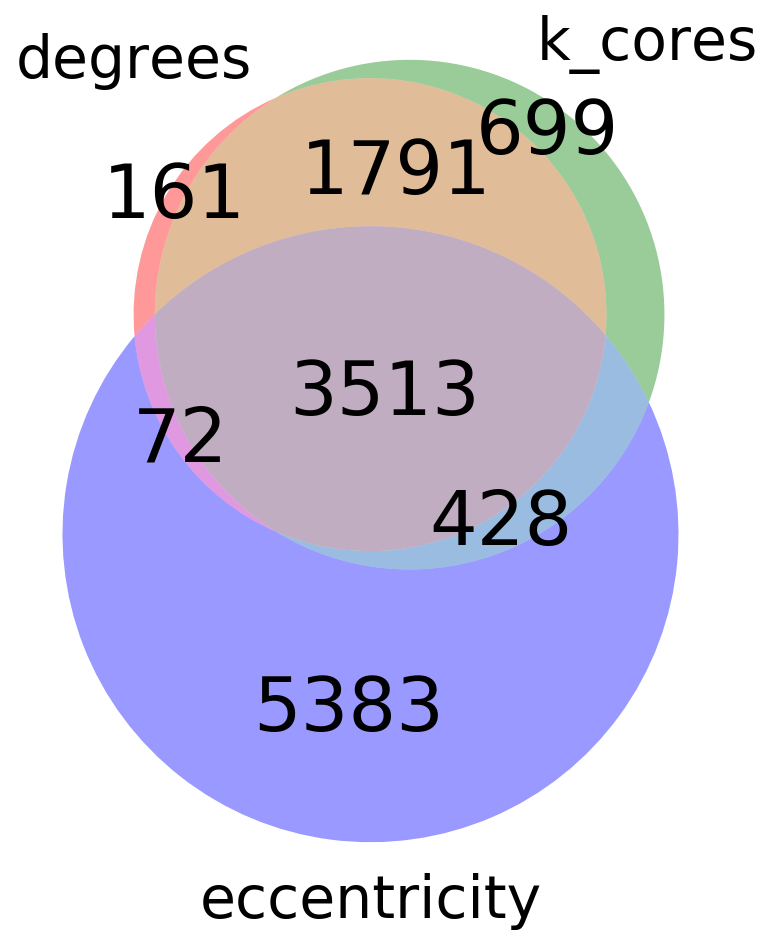}
\includegraphics[scale=0.5]{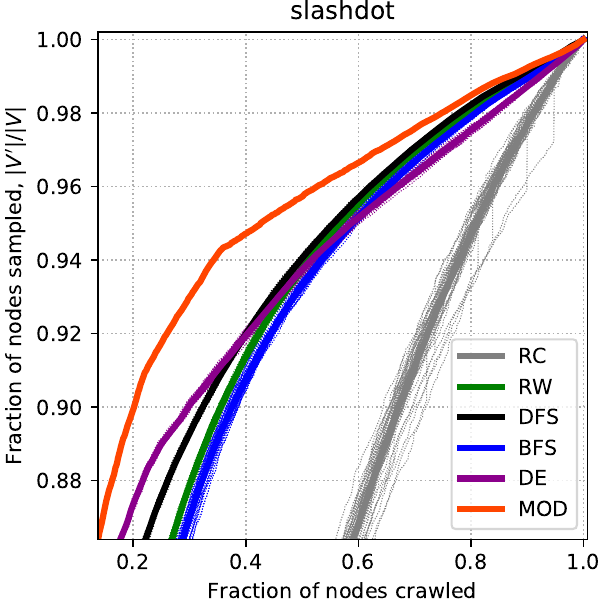}
% \caption{Venn diagram for top centrality nodes sets for slashdot graph}
\caption{Slashdot graph. Left: Venn diagram for top centrality nodes sets. Right: node coverage with variation for several seeds. Dotted liens correspond to individual seeds, bold lines correspond to averaged values.}
\label{fig:slashdot_venn_and_seeds}
\end{figure}

We take 10\% top-scored nodes of the graph as a target set in all our experiments.
Note that there are 4 different although overlapping target sets, one per each centrality measure. For example, Figure~\ref{fig:slashdot_venn_and_seeds} (left) shows 3 target sets from slashdot graph. The intersection of degree and k-coreness is significant, while the eccentricity set has about a half in common with them.

\begin{figure*}[h!]
\centering
\includegraphics[scale=0.5]{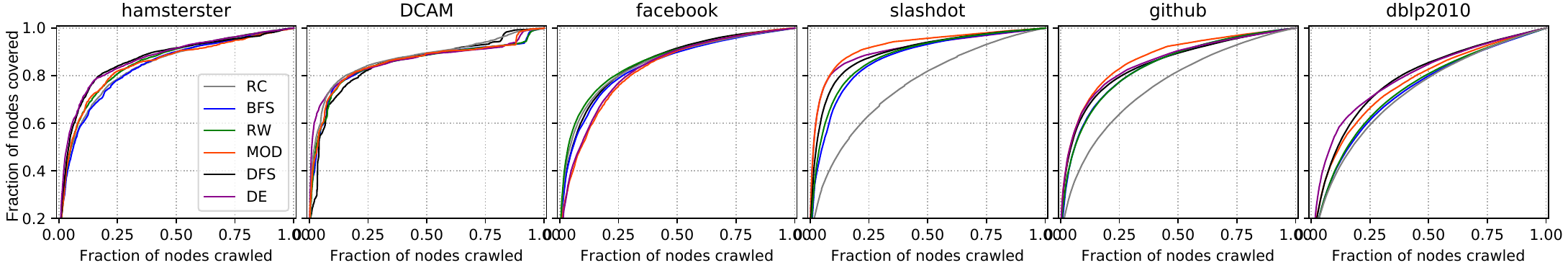}
\includegraphics[scale=0.5]{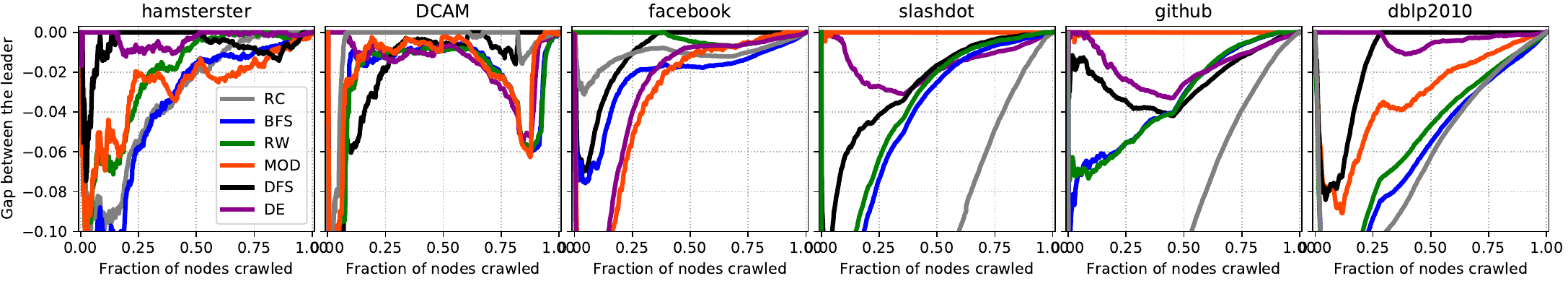}
\caption{Top: node coverage $c^{nodes}=|V'| / |V|$.
Bottom: same results, the gap between current method and the best result at each point (i.e. the higher the better) for node coverage.
}
\label{fig:all_graphs_nodes}
\end{figure*}

\begin{figure}[h!]
\includegraphics[scale=0.47]{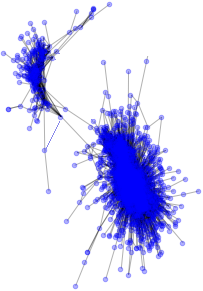}
\includegraphics[scale=0.5]{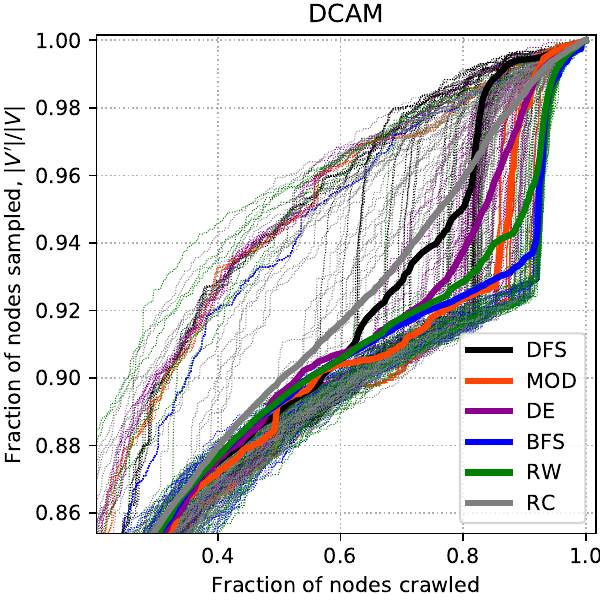}
\caption{DCAM graph. Left: network structure. Right: node coverage with variation for several seeds. Dotted liens correspond to individual seeds, bold lines correspond to averaged values.}
\label{fig:dcam_and_seeds}
\end{figure}

\subsection{Crawlers}
\label{sec:crawlers}

In our work, we considered 5 most popular crawling methods (RC, RW, DFS, BFS, MOD) and a recently proposed DE-Crawler. Table~\ref{tab:crawlers} summarizes the used algorithms together with their computational complexities per one iteration (in our implementations). 
Random Crawler (RC) and Random Walk (RW) algorithms each time go to a random node, selected from the whole observed set $V'_o$, or from the newly observed neighbours of the previously crawled node, respectively. RW is known to be effective at discovering top-centrality nodes~\cite{lim2011online}.
BFS and DFS implement two well-known search strategies. BFS crawler is usually applied for network analysis~\cite{mislove2007measurement}, although partial BFS crawls are biased towards high-degree nodes and underestimate low-degree nodes~\cite{lee2006statistical}.

MOD is a greedy method, based on a heuristic that a node~$v$ with high observed degree $deg(v, S)$ also has high real degree $deg(v, G)$. This proved to be efficient in terms of node coverage~\cite{avrachenkov2014pay, avrachenkov2012online}.

DE-Crawler was recently suggested by K.\,Areekijseree and S.\,Soundarajan~\cite{areekijseree2018crawler} as a smart combination of RW and a greedy algorithm. As it was shown in their previous work~\cite{areekijseree2018guidelines}, while MOD outperforms other methods, it gets stuck within communities in graphs with high modularity. At the same time, walking-based methods are good to move between dense regions of the network. %DE-Crawler combines the exploration strategy with expansion phase when it acts like a random walker.
DE-Crawler algorithm consists of two main stages: Densification and Expansion. These stages are switched, depending on the result of comparison of certain statistics. At each crawling step, $V'_o$ is sorted by $deg(v, S)$. In the Expansion mode, the next node to crawl is randomly selected from $80\%$ of bottom elements. In the Densificatoin mode, for the top $20\%$ a score $\Phi(v)$ is calculated: $\Phi(v) = \frac{deg(v, S)}{\langle deg(v, S) \rangle} (1-clust(v, S))$. The next node is the one with a maximal score, which is motivated by the observation that hubs have high degrees and low clustering coefficients~\cite{bloznelis2013degree}.

\begin{figure*}[h!]
\centering
\includegraphics[scale=0.49]{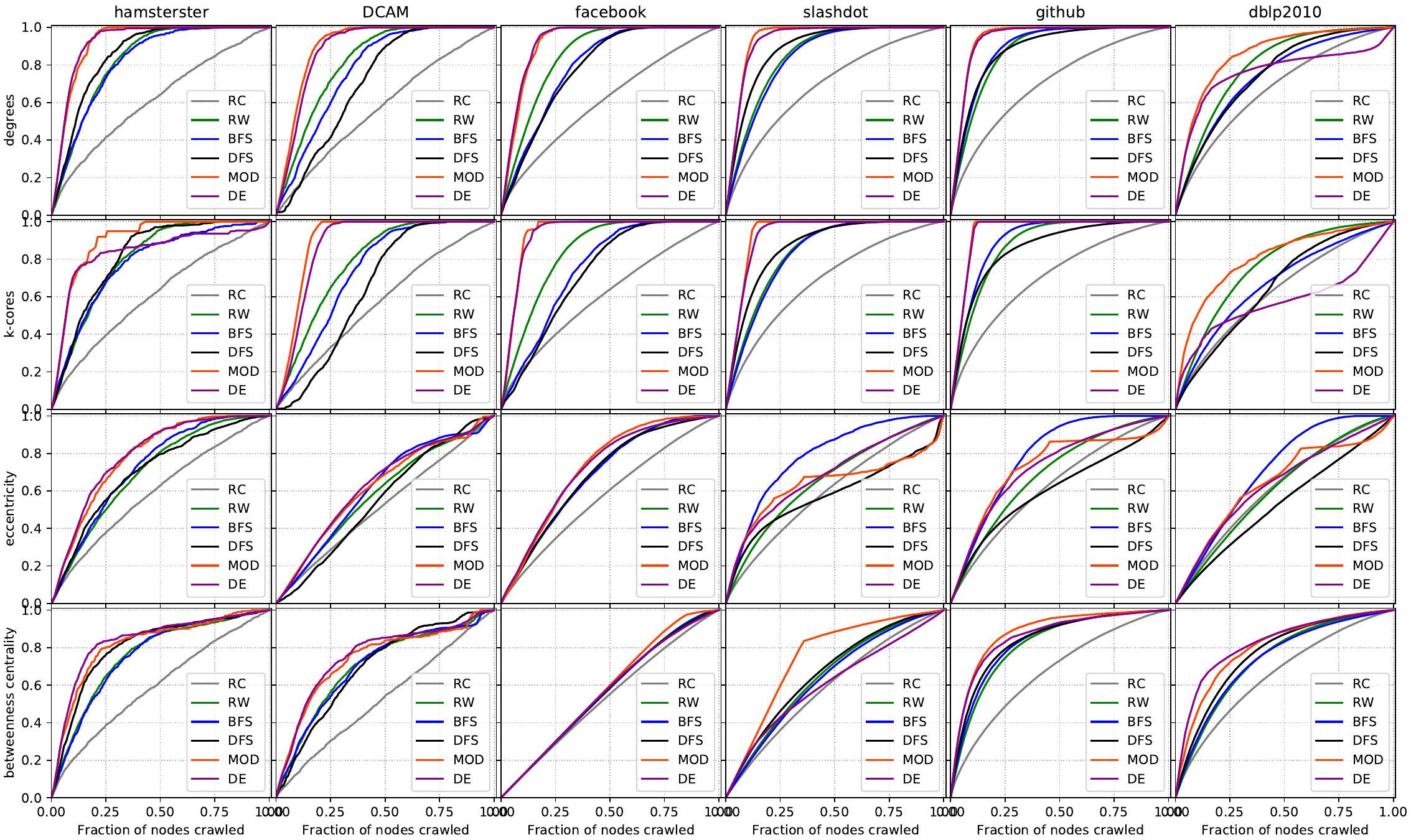}
\includegraphics[scale=0.48]{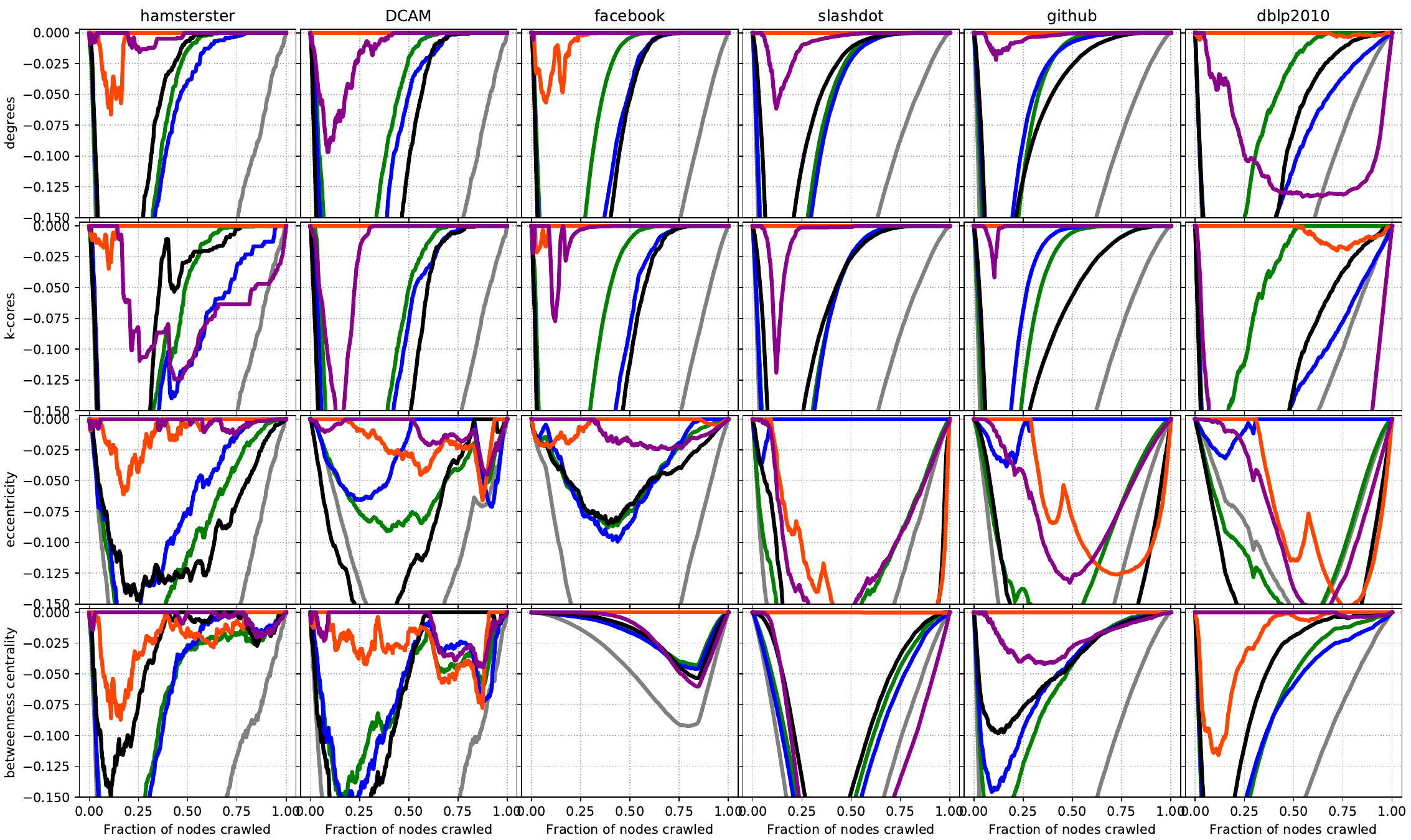}
\caption{Target set coverage $c^{target}_c=|V'_c \cap V^*| / |V^*|$ depending on the fraction of nodes crawled. The lower plot shows the gap between current method and the best result at each point (i.e. the higher the better).}
\label{fig:all_graphs_all_metrics_2}
\end{figure*}

\begin{figure*}[h!]
\centering
\includegraphics[scale=0.51]{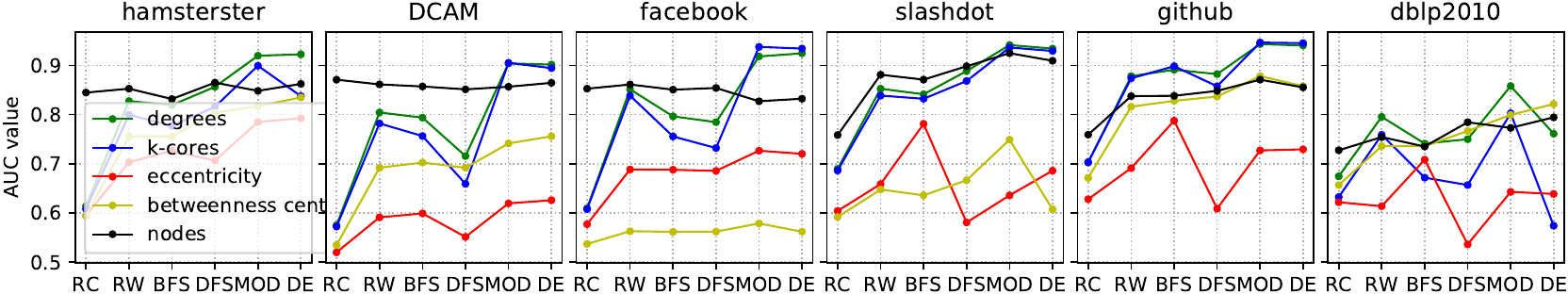}
\includegraphics[scale=0.51]{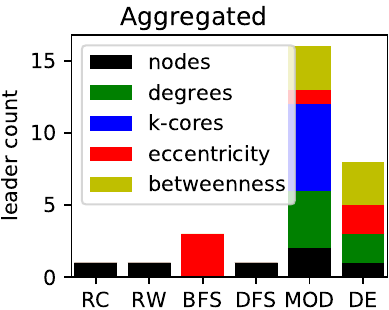}
\caption{AUC values computed for node coverage $c^{nodes}$ and target set coverage $c^{target}_c=|V'_c \cap V^*| / |V^*|$. Right: winners aggregated over the graphs. Colored bars' heights denote how many times this crawler was the best one at specific measures.}
\label{fig:all_graphs_AUC}
\end{figure*}

\subsection{Dataset}

For the experiments, we collected a dataset of small and medium-size networks from various domains. All graphs, except DCAM, are available at \textit{\url{http://networkrepository.com}}~\cite{nr} or at The Koblenz Network Collection \cite{kunegis2013konect}. DCAM was manually crawled from one community (\textit{\url{vk.com/club1694}}) by API and contains only open profiles. Since all considered crawlers are designed to operate with connected graphs only, we extracted a giant component from each graph and further analyzed it instead of the whole graph. Short descriptions and parameters of graphs can be found in Table~\ref{tab:graphs}.

\subsection{Method}

We tested the crawlers from Table~\ref{tab:crawlers} on all graphs from the dataset (Table~\ref{tab:graphs}). As it was mentioned in section~\ref{sec:problem}, a crawler starts with a randomly chosen seed node and traverses the graph, updating $V'_c$ and $V'_o$ sets. It stops when the whole graph is collected, i.\,e. $S=G$, $V'_c=V$, and $V'_o=\emptyset$.
In order to avoid bias of the seed choice, all results in the next section were averaged over 8 different seeds uniformly chosen from~$V$.

To evaluate the crawling algorithms, we used a classical measure, node coverage $c^{nodes}=|V'| / |V|$. We considered two ways to measure the coverage of a target set. One can count the coverage for all already known nodes, $c^{target}=|V' \cap V^*| / |V^*|$ or only for the closed ones: $c^{target}_c=|V'_c \cap V^*| / |V^*|$. The motivation for the second approach is that one is usually interested in how many influential nodes are collected rather than seen.

The target set $V^*$ is formed by the top $p=10\%$ of nodes sorted by one of four measures: node degree, betweenness centrality, k-coreness and eccentricity. For eccentricity we took the top nodes with the lowest value. For all cases we measured the coverage $c$ depending on the number of nodes crawled $i$, iterating $i$ from 1 to $|V|$.

\section{Experiments}
\label{sec:experiments}

For the experimental evaluation we implemented a framework with all six crawler methods: RC, RW, DFS, BFS, MOD, and DE. To the best of our knowledge, no public implementation is available, therefore we used our own. The computational complexities of one iteration of each algorithm are presented in Table~\ref{tab:crawlers}.

In all methods at each step we observe neighbours of the current node and add them to $V'_o$. This operation gives the lower bound of complexity $O(\langle deg(v, G) \rangle)$. %And the lower bound of total complexity $O(b*d_{avg})$. 

Random Crawler just takes a random node from $V'_o$ which costs O(1).
Random Walk crawler each time goes to a friend of the current node, which requires~O(1). However, it could surf among already closed nodes for an uncertain time until finds a node from $V'_o$. 

BFS and DFS algorithms require to keep the queue and stack, respectively. Getting a new node from those requires~$O(1)$.

The MOD method at each step requires a node from $V'_o$ with the maximal degree. We used sorted lists in which every node is sorted by its degree to speed  up our implementation. One step complexity is $ (2 \cdot deg(v, S) + 1) \cdot \log |V'_o| $.  So choosing the next node at each step became much more complex operation, because $deg(v, S)$ increases as size of $S$ increases.

Finally, DE is the most computationally complex algorithm. At each step, for each node in the top 20\% by degree in $|V_o'|$, it calculates statistics involving the clustering coefficient with complexity $ deg(v, S)^2 $. After that, it picks the node with the highest clustering coefficient as the next node. To improve the sorting step we also used sorted lists. The complexity to keep that list sorted at each step is $(2 \cdot deg(v, S) + 1) \cdot O(\log |V'_o|)$. The statistics require to find the average degree in $|V_o'|$ for calculating coefficients that decide on the mode switching. So, its total complexity is  $O(|V'_o|) \cdot (\langle deg(v, S)^2 \rangle + 1) + O(\log |V'_o|) \cdot (2 \cdot deg(v, S) + 1) \approx O(|V'_o| \cdot \langle deg(v, S)^2 \rangle) $, and it is much higher than that of all others.

\subsection{Nodes coverage}

We measured the node coverage $c^{nodes}=|V'| / |V|$ for all 6 crawler methods on 6 graphs. Results are presented at Figure~\ref{fig:all_graphs_nodes}.
The X-axis corresponds to the fraction of crawled nodes, from 0 to the 1 (the whole graph). The Y-axis shows a fraction of $|V'|$ to $|V|$.
We also plot the gap between current method and the best result at each point, i.e. the higher the better (bottom plot at Figure~\ref{fig:all_graphs_nodes}). Such a visualisation could help to see which method is the leader.
The main result is that the leader changes depending on the number of nodes crawled (budget size), for all graphs.
If the budget is limited to 5-10\% of $|V|$, DE crawler outperforms the others at hamsterster, DCAM, github, and dblp2010. But when the task is to collect the majority of the graph, DE crawler is not the optimal choice. For github and slashdot that would be MOD, while for dblp2010 and facebook --- DFS is the best choice. Results on small graphs hamsterster and DCAM are not so stable as on larger graphs: the leader changes several times depending on the budget.

Another observation concerns crawling curves on DCAM graph. When about 80-90\% of nodes are crawled, BFS, RW, and MOD curves demonstrate a hop (see Figure~\ref{fig:dcam_and_seeds}, right). The results are averaged over 50 random seeds; the averaged curve is plotted in bold, individual seeds are dotted lines.
The hopping behavior happens due to a specific structure of DCAM network. It consists of two well separated communities connected with a small bridge (Figure~\ref{fig:dcam_and_seeds}, left). MOD crawler is known to get stuck within communities with high modularity, while RW was reported to be able to transition between them~\cite{areekijseree2018guidelines}. However, we see that in average RW also gets stuck within the community.
DE crawler has a smoother curve and outperforms MOD, proving its ability to escape the community. Finally, according to this DCAM experiment, RC and DFS strategies are the best strategies for getting outside a dense community.

% seeds
\subsubsection{Seed choice influence}

Although an initial seed is expected to affect the crawling process, its influence is negligible at larger graphs. For example, at DCAM graph with 2.7K nodes, the seed choice could have a significant affect to crawler performance (see variability of dotted curves at Figure~\ref{fig:dcam_and_seeds}, right). 
But for slashdot graph with 51K nodes, results are already little dependent on the seed choice (see Figure~\ref{fig:slashdot_venn_and_seeds}, right). 
Nevertheless, it should be noted that for slashdot at early crawling stage, when budget is comparable to the size of DCAM, a similar high variability is observed.
The same holds for larger graphs from the dataset.

\subsection{Influential nodes coverage}

We measured the target set coverage for 6 crawler methods on 6 graphs of our dataset. The results of the node coverage are presented in Figure~\ref{fig:all_graphs_nodes}, the target set coverage for 4 measures is shown at Figure~\ref{fig:all_graphs_all_metrics_2} 
(top). 
We show only $c^{target}_c$ measure results since it reflects a more realistic picture of collecting most influential nodes. The other reason is that for $c^{target}$ measure, method lines goes too close to each other on the plots to distinguish them.

The lower plot of Figure~\ref{fig:all_graphs_all_metrics_2} shows the gap between current method and the best result at each point, i.e. the higher the better.

The first thing one can see is that the results vary a lot for different measures at different graphs.
RC often resembles a straight line but is convex at bigger graphs. This means that even such a simple strategy of randomly selecting one of the observed nodes works better than if we would pick a random node of the graph (which would give a straight line on the plot).

\subsubsection{Degree and k-coreness}

For collecting top-degree nodes, as one could expect, greedy algorithms MOD and DE are better than the others. Comparing MOD and DE (see lower plot of Figure~\ref{fig:all_graphs_all_metrics_2}), one can see that DE does not outperform MOD. Moreover, MOD outperforms DE at slashdot, github, and dblp2010, while DE is strongly better only at facebook.

Quite a similar picture one can see for k-cores centrality at first 5 graphs. For both centralities, MOD and DE are leaders, RC performs the worst, while RW, BFS, and DFS are in the middle almost all the time.

A surprising behavior one can see at dblp2010, where DE is much worse.
We assume that the reasons are the unusual topological structure (connected stars) of the dblp2010 graph and the default coefficients in DE statistics formulas. Switching between densification and expansion modes did not occur at the moment when it was intended. It is likely that a more accurate tuning of these coefficients is needed to make DE-crawler behave as expected by its authors.

\subsubsection{Eccentricity}

A different result we observed for eccentricity measure. For slashdot, github, and dblp2010, the BFS strategy significantly outperforms the others. Moreover, MOD loses its leading position after 10-20\% of steps and then loses all other methods. 

Interestingly, its coverage curve is not smooth and have several turning points. 
We believe MOD eventually gets stuck within communities composed of vertices with high eccentricity. 
For any vertex, difference of its eccentricity and eccentricities of neighbouring vertices does not exceed one.
Therefore, if a vertex has high eccentricity, its neighbours also have high eccentricity, and vice versa, if a vertex belongs to target set of vertices with lowest eccentricity, its neighbours probably belong there too.
Thus there are communities with lots of vertices from target set and there are communities with none of them.
When MOD gets stuck in community with high eccentricity, its coverage curve goes flat.
Finally, for these three graphs the DFS algorithm also works bad, even worse than RC for most of the time.

\subsubsection{Aggregated results}

We also computed area under curve (AUC) for the results mentioned before, in order to compare crawlers performance independently of budget (Figure~\ref{fig:all_graphs_AUC}, left).
One can see that in most cases low eccentricity nodes are detected worse than other centralities. 
Results for top degree nodes and top k-core nodes correlate well, since these measures are related. This is also consistent with Venn diagram (Figure~\ref{fig:slashdot_venn_and_seeds}, left) showing a significant overlap between the corresponding sets of nodes.

We aggregated AUC results over all graphs, counting how many times each crawler was the best one for a specific measure (Figure~\ref{fig:all_graphs_AUC}, right). MOD appeared to get the highest score of 16, two times more than DE. MOD is also the best for k-core nodes collecting for all 6 graphs.
BFS is good at lowest eccentricity nodes crawling, possibly due to very bad results of MOD and DE on slashdot, github, and dblp2010 graphs. However, this observation needs a more detailed analysis.
Finally, note that results on node coverage are the most uncertain: 5 of 6 methods were the winner at least once. 

\section{Related work}
\label{sec:related}

There are many studies in literature on the network crawling problem. For a detailed review, please refer to a survey~\cite{ahmed2014network}.

Ye Shaozhi and co-authors in~\cite{ye2010crawling} investigate the network crawling problem on four social networks: Orkut, Youtube, Live Journal and Flickr. They analyze several crawlers depending on seed choice, network structure, and presence of protected profiles.
Among their findings are the following ones. A small number of steps is enough to obtain a large part of the target graph: 10\% of nodes being crawled provide 49--74\% of nodes of the graph. Node/link coverage does not depend much on the choice of seeds: by choosing high degree seeds the improvement did not exceed 5\%. Greedy algorithms, like MOD, achieve high node/link coverage faster, but are less robust than BFS.

Several works aim at sampling top centrality nodes.
Yeon-sup Lim et al. in their work~\cite{lim2011online} try to estimate nodes with top-$k$ centrality, namely degree, betweenness, and closeness. Experiments on various networks show that RW crawler quickly discovers a major fraction of the top-degree nodes: 5\% of the sampled nodes contain over 80\% of the top-10. Node degree correlates well with other centralities and thus could be used as an approximation for them. For betweenness and closeness measures, RW strategy outperforms more complex strategies.
Two kinds of error are introduced, sampling (collection) error and identification error. A sampling error means that a node from the top-$k$ is missing in the sample, while an identification error occurs when the correct node is sampled but not recognized as such.
The authors discover that sampling error is higher when the network has slightly skewed degree distribution. When the degree centrality has low correlation to betweenness and closeness in a network, identification error become larger.

A challenging task could be to detect top-$k$ centrality nodes in a known large graph, for example, in a crawled sample. Since the calculation of betweenness and closeness centralities requires search for all-pairs shortest paths, which complexity is $O(|V||E|)$ for an undirected graph, their exact computation could be too expensive~\cite{kang2011centralities}.
A series of works employ compressive sensing approach to detect top centrality nodes, e.g.~\cite{mahyar2015detection,mahyar2018identifying,mahyar2018compressive,mahyar2019compressive}. The idea is to use a part of the known graph for estimations based on compressive sensing theory.
K.\,Avrachenkov, N.\,Litvak, and coauthors suggested a sublinear  algorithm to finding list of nodes with maximal degree~\cite{avrachenkov2014quick, avrachenkov2010monte}. A random walk based approach allows to approximately find such nodes faster than in $O(|V|)$ iterations.
Re-weighted random walks are used in order to achieve uniform crawling~\cite{gjoka2011practical}.

In addition to the six crawlers considered in our paper, alternative algorithms were suggested in literature. Maximum Observed PageRank method~\cite{salehi2012sampling}, Volatile Multi-armed Bandit~\cite{bnaya2013bandit}, parallel crawlers~\cite{chau2007parallel}, Online Page Importance Computation~\cite{abiteboul2003adaptive}.

\section{Conclusion}
\label{sec:conclusion}

We tested six network crawlers on several graphs from various domains. 5 crawlers are well known, RC, RW, BFS, DFS, MOD, and 1 more DE-crawler was proposed recently. We run the algorithms until they collect the whole graph and measured the coverage of a target set of top-10\% influential nodes.
We obtained the following results.

There is no superior crawling method neither for classical nodes coverage measure, nor for influential nodes coverage. Furthermore, the leading algorithm for a particular graph can change several times depending on the fraction of graph nodes is collected. MOD, DE, RW, BFS, and DFS methods all have been the leader for several times in our experiments.
This is in contrast with result obtained by running crawlers with a limited budget, e.g.~\cite{areekijseree2018guidelines,areekijseree2018crawler}. Our results on a whole graph imply that the leading crawler do change depending on the budget.

Greedy strategy MOD is often the optimal one, but not always, which is consistent with known results from literature.
This is also true for collecting of influential nodes. 
%There are cases when MOD loses to all other considered methods. 
For example, at slashdot, github, and dblp2010 graphs, MOD outperforms the others at lowest eccentricity nodes crawling at early stages (until 10-25\% of nodes of graph are collected). But further it loses to all other considered methods.
Surprisingly, BFS showed the best coverage of lowest eccentricity nodes.

The previous conclusion also holds for DE crawler.
Moreover, experiments do not evidence that DE outperforms MOD, there are cases when it loses to many other methods. Taking into consideration its high computational complexity, it is not currently a good choice. 
Probably, a more accurate parameters tuning could improve its performance.

Concerning the task of collecting centrality measures, MOD is the best one for k-coreness. MOD and DE are better than the other 4 methods in collecting top degree and betweenness centrality nodes. As for lowest eccentricity nodes, the best method could not be reliably defined. All 6 crawlers detect this kind of nodes worse (relatively slower) than the other types of influential nodes.

In one experiment with DCAM graph consisting of two well separated communities, we observed that RC and DFS are less prone to getting stuck within highly modular communities, than RW, MOD, DE, and BFS.

The influence of the seed choice is not crucial for comparing crawlers, when graph size is more than 50K nodes. This confirms the results of Ye Shaozhi et al~\cite{ye2010crawling}.

\bibliography{main}
\bibliographystyle{IEEEtran}

\end{document}